\documentclass[10pt, conference, compsocconf]{IEEEtran}
\IEEEoverridecommandlockouts

\usepackage{array}
\usepackage{cite}
\usepackage{url}
\usepackage{graphicx}
\usepackage{balance}  %
\usepackage{soul}
\usepackage{graphics}
\usepackage{mdwlist}
\usepackage[ruled, noend, linesnumbered]{algorithm2e} %

\usepackage{booktabs}
\usepackage{soul}
\usepackage[center]{subfigure}

\usepackage{amsmath}
\usepackage{amsfonts} 
\usepackage{amsbsy}
\usepackage{amsthm}

\usepackage[]{caption}
\usepackage{multirow}
\usepackage{mathrsfs}
\usepackage[mathscr]{eucal} %
\usepackage{xspace}
\usepackage{paralist} %
\usepackage{mathtools}
\usepackage{verbatim}
\usepackage{tabularx}
\usepackage{makecell}
\usepackage{xcolor}
\usepackage{comment}
\usepackage{enumitem}
\usepackage{pifont}

\newtheorem*{proof}{Proof}

\newcommand{\smallsection}[1]{{\vspace{0.3mm}\noindent {\bf{\underline{\smash{#1}}}}}}

\newcommand{\but}[1]{\underline{\textbf{\smash{#1}}}}
\newcommand{\ut}[1]{\underline{\smash{#1}}}

\definecolor{mygreen}{RGB}{0, 100, 0}

\newcommand{\red}{\textcolor{red}}

\newcommand{\htails}{heavy-tailed distributions\xspace}
\newcommand{\expdist}{exponential distribution\xspace}

\newcommand{\Lcal}{\mathcal{L}}
\newcommand{\Dcal}{\mathcal{D}}
\newcommand{\model}{\textsc{HyperFF}\xspace}
\newcommand{\hyperpa}{\textsc{HyperPA}\xspace}

\newcommand{\edge}{e_t}
\newcommand{\edges}{\edge^s}
\newcommand{\edgesseq}{\{\edges\}_{t=1}^T}
\newcommand{\nulledge}{e_t'}
\newcommand{\nulledges}{(\nulledge)^s}
\newcommand{\nulledgesseq}{\{\nulledges\}_{t=1}^T}
\newcommand{\graphseq}{\{G_t\}_{t=1}^T}
\newcommand{\Vt}{V_t}

\newcommand{\addgets}{\xleftarrow[]{\text{add}}}
\newcommand{\setgets}{\xleftarrow[]{\text{set}}}

\let\oldnl\nl%
\newcommand{\nonl}{\renewcommand{\nl}{\let\nl\oldnl}}%
\newcolumntype{x}[1]{>{\centering\arraybackslash\hspace{0pt}}p{#1}}

\newcommand{\contactl}{Contact\xspace}
\newcommand{\contact}{contact\xspace}
\newcommand{\email}{email\xspace}
\newcommand{\tags}{tags\xspace}
\newcommand{\ndc}{substances\xspace}
\newcommand{\thread}{threads\xspace}
\newcommand{\coauth}{coauth\xspace}

\begin{document}

\title{Evolution of Real-world Hypergraphs: Patterns and Models without Oracles}

\author{\IEEEauthorblockN{Yunbum Kook}
\IEEEauthorblockA{Dept. of Mathematical Sciences, KAIST \\
yb.kook@kaist.ac.kr}
\and
\IEEEauthorblockN{Jihoon Ko}
\IEEEauthorblockA{Graduate School of AI, KAIST \\
jihoonko@kaist.ac.kr}
\and
\IEEEauthorblockN{Kijung Shin}
\IEEEauthorblockA{Graduate School of AI \& School of EE, KAIST \\
kijungs@kaist.ac.kr}
}

\maketitle

\begin{abstract}
	What kind of macroscopic structural and dynamical patterns can we observe in real-world hypergraphs?
What can be underlying local dynamics on individuals, which ultimately lead to the observed patterns, beyond apparently random evolution?

Graphs, which provide effective ways to represent pairwise interactions among entities, fail to represent group interactions (e.g., collaboration of three or more researchers, etc.).
Regarded as a generalization of graphs, hypergraphs allowing for various sizes of edges prove fruitful in addressing this limitation.
The increased complexity, however, makes it challenging to understand hypergraphs as thoroughly as graphs.

In this work, we closely examine seven structural and dynamical properties of real hypergraphs from six domains.
To this end, we define new measures, extend notions of common graph properties to hypergraphs, and assess the significance of observed patterns by comparison with a null model and statistical tests.

We also propose \model, a stochastic model for generating realistic hypergraphs.
Its merits are three-fold: 
(a) \ut{Realistic:} it successfully reproduces all seven patterns, in addition to five patterns established in previous studies,
(b) \ut{Self-contained:} unlike previously proposed models, it does not rely on oracles (i.e., unexplainable external information) at all, and it is parameterized by just two scalars,
and (c) \ut{Emergent:} it relies on simple and interpretable mechanisms on individual entities, which do not trivially enforce but surprisingly lead to macroscopic properties.

\end{abstract}

\section{Introduction}
\label{sec:intro}

Which structural patterns do real-world hypergraphs have and how do they evolve over time?
Are there simple mechanisms on individual nodes that these patterns emerge from?

Datasets based on relationships between two objects have naturally arisen from a wide range of domains in the real world: friendships between two users in online social networks, hyperlinks from a web page to another, and citations from a publication to another, to name a few.
Graph representations give rise to an easy and extensive analysis of this type of relational datasets, and they have been used to understand such datasets in many respects.

A thorough understanding of those datasets via graph analysis has resulted in insights behind them and facilitated the development of effective algorithms building on the insights.
Some of the well-known static properties are a small-world phenomenon, also known as six degrees of separation \cite{travers1977experimental}, and power law distributions of spectra \cite{farkas2001spectra} and degrees \cite{faloutsos1999power} of graphs in various domains.
Temporal properties prevalent in real graphs include triadic closure \cite{huang2014mining}, temporal locality \cite{shin2017wrs}, densification, and shrinking diameter over time \cite{leskovec2007graph}.
These informative properties actually serve as useful tools for designing and analyzing graph algorithms \cite{cooper2012fast, latapy2008main, gleich2012vertex, ko2020mosso}.

Previous studies on fundamental mechanisms leading to these patterns have also proceeded with the development of generative models for realistic graphs.
Various models \cite{kleinberg1999web, newman2001clustering, leskovec2007graph, leskovec2010kronecker} have been proposed to reproduce the examined patterns and lent themselves to some applications, such as simulations \cite{vinciguerra2010geography, kaiser2007simulation}, benchmarking \cite{edunov2016darwini}, and sampling \cite{leskovec2006sampling}.

\begin{figure}[t]
	\centering
	\hspace{-2mm}
	\includegraphics[width=0.75\columnwidth]{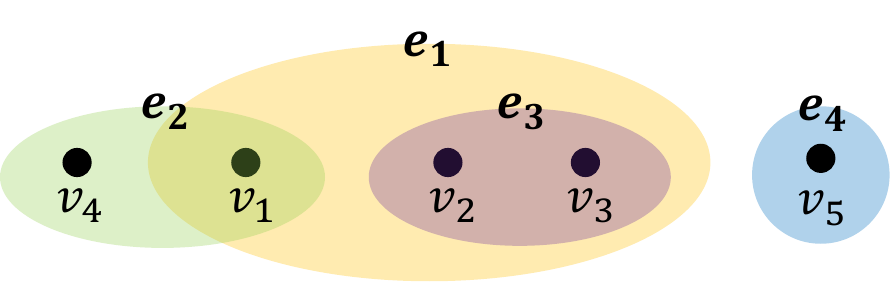}
	\caption{\label{fig:example}
		\but{An example hypergraph} with five nodes ($v_{1}$, $v_{2}$, $v_{3}$, $v_{4}$, and $v_{5}$) and four hyperedges ($e_{1}$, $e_{2}$, $e_{3}$, and $e_{4}$). Note that different hyperedges may contain different numbers of nodes.
	}
	\vspace{-6mm}
\end{figure}

\begin{figure*}[t]
	\centering
	\caption{\label{fig:ad} 
		\but{\model generates realistic hypergraphs}.
		In the first row, we observe the heavy-tailed distributions of \textbf{(S1)} degrees, \textbf{(S2)} hyperedge sizes, \textbf{(S3)} intersection sizes, and \textbf{(S4)} singular values of incidence matrices; and \textbf{(T1)} diminishing overlaps of hyperedges, \textbf{(T2)} increasing edge density (more clear in the other datasets), and \textbf{(T3)} shrinking diameter in real-world hypergraphs.
		In the second row,
		\model successfully reproduces all the seven patterns. See Sections~\ref{sec:patterns} and \ref{sec:model} for details.
	}
	\scalebox{0.82}{
		\begin{tabular}{c|ccccccc}
			\toprule
			\multicolumn{1}{c}{} &
			\multicolumn{4}{c}{\textsc{Structural Patterns}} &
			\multicolumn{3}{c}{\textsc{Dynamic Patterns}}	
			\\
			\cmidrule(lr){2-5}
			\cmidrule(lr){6-8}
			\multicolumn{1}{c}{} &
			Degrees &
			Hyperedge Sizes &
			Intersection Sizes &
			Singular Values &
			Intersecting Pairs &
			Edge Density &
			Diameter
			\\
			\cmidrule(lr){2-5}
			\cmidrule(lr){6-8}
			\rotatebox[origin=l]{90}{Real Data}
			&
			\includegraphics[height=0.762in]{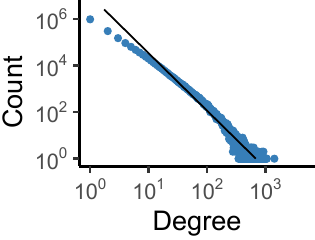} &
			\includegraphics[height=0.762in]{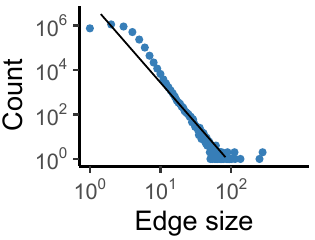} &  	
			\includegraphics[height=0.762in]{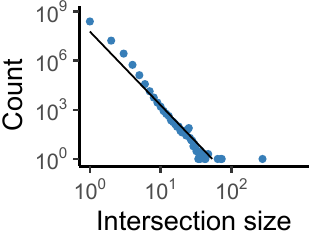} &
			\includegraphics[height=0.762in]{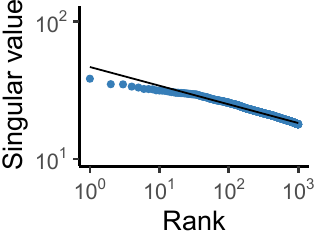} &
			\includegraphics[height=0.762in]{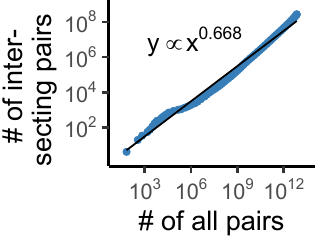} &
			\includegraphics[height=0.762in]{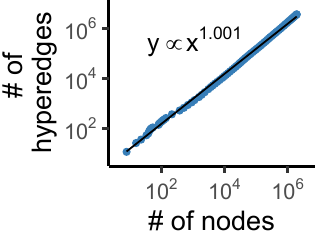} &
			\includegraphics[height=0.762in]{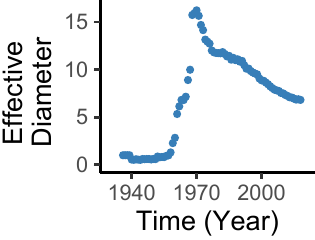}
			\\
			\cmidrule(lr){2-5}
			\cmidrule(lr){6-8}
			\rotatebox[origin=l]{90}{\small \textbf{\model}}
			\rotatebox[origin=l]{90}{\small \textbf{(Proposed)}}
			& 
			\includegraphics[height=0.762in]{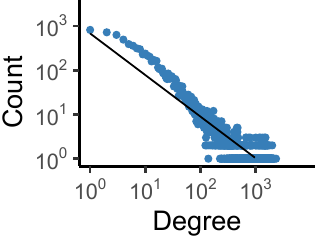} &
			\includegraphics[height=0.762in]{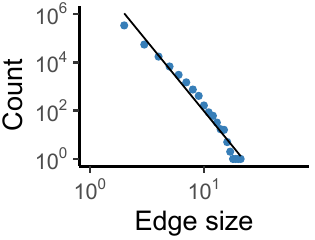} &  	 	 			
			\includegraphics[height=0.762in]{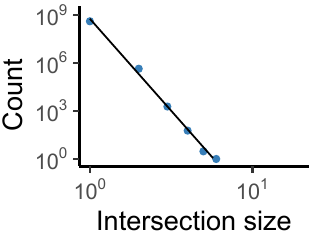} &
			\includegraphics[height=0.762in]{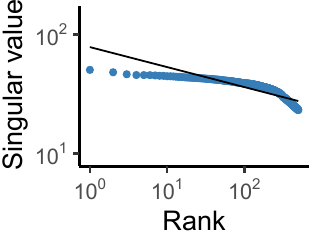} &
			\includegraphics[height=0.762in]{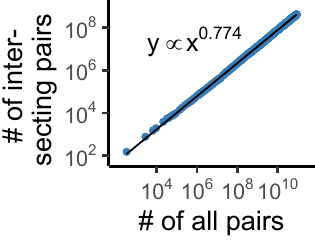} &
			\includegraphics[height=0.762in]{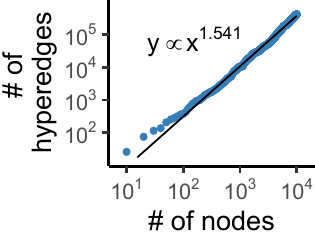} &
			\includegraphics[height=0.762in]{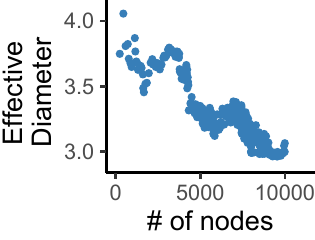} \\
			\bottomrule
		\end{tabular}
	}
	\vspace{-5mm}
\end{figure*}

Not all relational networks are restricted to have only pairwise relationships, and polyadic relationships are also ubiquitous \cite{benson2016higher,wasserman1994social,temkin1996chemical,pimm1982food,benson2018sequences};
for instance, publishing a paper may involve three or more authors, and an online group chat may involve tens of participants.

\emph{Hypergraphs} \cite{berge1984hypergraphs, berge1973graphs,chodrow2020annotated} are a natural extension of the conventional notion of graphs by allowing for various sizes of edges.
Formally, a hypergraph consists of a set of \emph{nodes} and a set of \emph{hyperedges}, where each hyperedge is a non-empty subset containing any number of nodes.
They can represent high-order interactions (i.e., interactions among any number of objects), not just between two as in the graphs.

Taking into account polyadic interactions proves fruitful and indeed inevitable for challenging tasks.
According to \cite{yoon2020much}, the harder tasks are, the more useful leveraging information on high-order interactions is, and a simple reduction from high-order to pairwise interactions may lead to significant performance degradation.
Hence, hypergraphs, which naturally represent such complex interplays, have drawn considerable attention in many domains, including  computer vision \cite{huang2009video,huang2010image,wang2015visual}, recommendation \cite{bu2010music, li2013news, tan2011using}, and graph learning \cite{zhou2007learning, yu2012adaptive, zhang2018dynamic, jiang2019dynamic,feng2019hypergraph, yadati2019hypergcn, yadati2018link}.

Despite its importance, the understanding of real hypergraphs is still not as concrete as that of real graphs.
The daunting complexity due to the variability in hyperedge sizes prevents straightforward extensions of concepts and tools used for graphs to hypergraphs.
Nevertheless, there have been fruitful and insightful attempts.
In \cite{yoon2020much, manh2020multi}, hypergraphs are decomposed into conventional graphs, and well-known properties of graphs on the decomposed graphs are examined.
In \cite{benson2018simplicial}, the triadic closure theory is extended to hypergraphs, and in \cite{benson2018sequences}, repetitive patterns of hyperedges are investigated in terms of subset correlation and recency bias.

Driven by the importance of hypergraphs and of extensive studies, we scrutinize additional four structural and three dynamical patterns inherent in real-world hypergraphs at the macroscopic level.
To this end, we come up with measures to capture new aspects, revisit well-known properties of ordinary graphs, and study their hypergraph analogs.
We thoroughly validate the significance of each observed pattern by relying on qualitative and quantitative analyses.

The established structural and dynamical patterns constitute a meaningful step toward advancing the partial understanding of real-world hypergraphs.
For structural patterns, we suggest that distributions of \textbf{(S1)} degrees
, \textbf{(S2)} hyperedge sizes, \textbf{(S3)} intersection sizes of hyperedges, and \textbf{(S4)} singular values of incidence matrices
fall under the class of heavy-tailed distributions, of which the first three and the last are close to a truncated power law and log-normal distribution, respectively, among probable candidates.
For dynamical patterns, we observe that, over time, \textbf{(T1)} the overlapping of hyperedges becomes less frequent,
\textbf{(T2)} the number of hyperedges grows faster than that of nodes (i.e., densification), and \textbf{(T3)} the distances between nodes decrease (i.e., shrinking diameter). 

Using these patterns as criteria, we propose a stochastic model \model (\textbf{Hyper}graph \textbf{F}orest \textbf{F}ire) for realistic hypergraph generation.
In \model, where a hypergraph grows with new nodes, each new node goes through two stages.
In the first stage, a node is randomly chosen and a `forest fire' starts there.
The forest fire is spread through existing hyperedges stochastically, and the new node forms a hyperedge with each of the burned nodes.
In the second stage, each new hyperedge expands similarly through a forest fire.

The benefits of \model are three-fold.
First, \model successfully reproduces all seven observed patterns, whereas previously proposed models focus only on a narrow scope of patterns.
We also apply the decomposition technique \cite{manh2020multi} to generated hypergraphs and confirm the five known properties of decomposed graphs.
Second, while the previous models rely on unexplainable external information (e.g., the number of new hyperedges along with each new node \cite{manh2020multi} and the size of each new hyperedge \cite{manh2020multi,benson2018sequences} with the number of new nodes in it \cite{benson2018sequences}),
\model requires no such oracles, being parameterized simply by two scalars: burning and expanding probabilities.
Lastly, \model leads to a deeper understanding of complex systems, giving a simple underlying mechanism that imposes the non-trivial macroscopic patterns.

As in the exploration of real hypergraphs, we validate \model through quantitative and qualitative analyses.
We also explore its parameter space and suggest values of parameters for generating realistic hypergraphs.

Our contributions are summarized as follows:
\begin{itemize}[leftmargin=*]
	\item \textbf{Establishment of structural and dynamical patterns in real-world hypergraphs}.
	\begin{enumerate}
		\item \textbf{Structural:} Heavy-tailed distributions of degrees, hyperedge sizes, intersection sizes, and singular values of incidence matrices.
		\item \textbf{Dynamical:} Diminishing overlaps of hyperedges, densification, and shrinking diameter.
	\end{enumerate}
	\item \textbf{Stochastic model \model for hypergraph generation}.
	\begin{enumerate}
		\item \textbf{Realistic:} It exhibits all seven observed patterns and the five structural patterns reported in the previous study.
		\item \textbf{Self-contained:} It does not rely on oracles or external information, and it is parameterized by just two scalars.
		\item \textbf{Emergent:} Its simple and interpretable mechanisms on individual nodes non-trivially produce the examined patterns at the macroscopic level.
	\end{enumerate}
\end{itemize}
\smallsection{Reproducibility:} We make the source code and datasets used in this work publicly available at \url{https://git.io/JUfAr}.

In Section~\ref{sec:related}, we survey a line of research on
properties and generative models of real-world (hyper)graphs.
In Section~\ref{sec:prelim}, we introduce some notations and concepts.
In Section~\ref{sec:patterns}, we examine the structural and dynamical patterns in real-world hypergraphs.
In Section~\ref{sec:model}, we propose our model \model and validate it.
We conclude our work in Section~\ref{sec:conclusions}.

\section{Related Work}
\label{sec:related}

Below, we review previous work on patterns in real-world hypergraphs and generation of realistic (hyper)graphs.

\subsection{Patterns in Real-world Hypergraphs.}
There have been considerable efforts to find out fundamental properties in the structures and dynamics of real-world hypergraphs.
Structural patterns are investigated in static hypergraphs or a few snapshots of evolving hypergraphs. Dynamical patterns are investigated in time-evolving hypergraphs.

Instead of dealing directly with the complexity of hypergraphs, \cite{yoon2020much, manh2020multi, benson2018simplicial} reduce  hypergraphs to a set of ordinary graphs, which contains partial or all information on the hypergraphs, and then identify the properties of the graphs, using well-defined graph measures.
The most basic way is to convert a hypergraph into the graph with the same set of nodes and the set of edges induced by replacing each hyperedge with a clique \cite{benson2018simplicial}, referred to as the $2$\emph{-projected graph}.
On $2$-projected graphs, \cite{benson2018simplicial} investigates simplicial closure, the extension of triadic closure to hypergraphs, and \cite{estrada2006subgraph} defines sub-hypergraph centrality and extends the notion of clustering coefficients to hypergraphs.
This simple abstraction is extended to $n$\emph{-projected graphs} \cite{yoon2020much} and $n$\emph{-level decomposed graphs} \cite{manh2020multi}, which aim to capture the interactions between the subsets of $n$ nodes in distinct ways.
Especially for the $n$-level decomposed graphs, we elaborate on their definition in Section~\ref{sec:model:exp}.
Based on these alternatives, \cite{manh2020multi} shows that the $n$-level decomposed graphs obtained from real hypergraphs retain five structural patterns: giant connected components, heavy-tailed degree distributions, small diameters, high clustering coefficients, and skewed singular-value distributions.
Besides these approaches,
\cite{lee2020hypergraph} focuses on local structural patterns, specifically the patterns of interactions among three hyperedges; and
\cite{benson2018sequences} focuses on dynamical patterns regarding the subset correlation, recency bias, and repeat patterns of hyperedges.
More patterns concerning diffusion, synchronization, and games are surveyed in \cite{battiston2020networks}.
In this work, we study seven additional generic patterns in real hypergraphs at the macroscopic level.

\subsection{Generation of Realistic (Hyper)graphs.}
A number of generative models have been proposed for realistic graph generation.
For example, preferential attachment \cite{cooper2003general,barabasi1999emergence} and copying models \cite{kleinberg1999web} provide simple underlying principles leading to the power law in degree distributions.
The forest fire model \cite{leskovec2007graph} reproduces two dynamical patterns in graphs: densification (i.e., increase in the number of edges per node) and shrinking diameter (i.e., decrease in the the distances between nodes) over time. 
The key idea behind the model is to connect a new node with a random node and the nodes burned by a `forest fire', which starts at the random node and is spread stochastically along the edges.
The Kronecker model \cite{leskovec2010kronecker} is a tractable model that enjoys a well-developed theory for the Kronecker product and thus provides theoretical guarantees as well as reproduce several established patterns.
Darwini \cite{edunov2016darwini} aims to generate realistic large-scale graphs whose distributions of degrees and clustering coefficients fit explicitly given distributions.
GraphRNN \cite{you2018graphrnn} takes a machine learning approach to train an autoregressive model generating graphs structurally similar to a given set of graphs.

Compared to realistic graph generation, relatively little attention has been paid to realistic hypergraph generation.
The stochastic model proposed by \cite{benson2018sequences} takes into consideration recurrent patterns of hyperedges;
however, it requires an oracle which provides the sizes of the next hyperedges, and the number of new nodes in them.
The generative model suggested in \cite{manh2020multi} extends the preferential attachment model to hypergraphs. In the model called \hyperpa, for each new node, the group of nodes forming a hyperedge with the new node is chosen randomly, proportionally to the number of existing hyperedges that contain the group.
While \hyperpa generates a hypergraph whose $n$-decomposed graphs exhibit the five empirical structural patterns, it requires in advance the distribution of hyperedge sizes and the number of hyperedges related to each new node.
However, our model described in Section~\ref{sec:model} is able to reproduce the seven patterns examined in this paper without any oracles, and its decomposed graphs also maintain the five structural patterns.

\section{Preliminaries}
\label{sec:prelim}

\begin{figure*}[t]
	\centering
	\caption{\label{fig:static:patterns}
		\but{Structural properties shown in six real datasets: Heavy-tailed distributions of four quantities}.
		The first three properties mostly show well-suited straight lines on the log-log scale, implying the potential for a stronger claim: they follow a power law distribution.
		This tendency is more apparent in larger hypergraphs.
		See Section~\ref{sec:patterns:static} for details.
	}
	\scalebox{0.97}{
		\hspace{-3mm}
		\begin{tabular}{p{0.024\textwidth}|cccccc}
			\toprule
			{\footnotesize Data~} & \contact (smallest) & \email & \tags & \ndc & \thread & \coauth (largest)\\ 
			\midrule
			\midrule
			\rotatebox[origin=l]{90}{\small $ $ $ $ $ $ Degrees}
			&
			\includegraphics[height=0.762in]{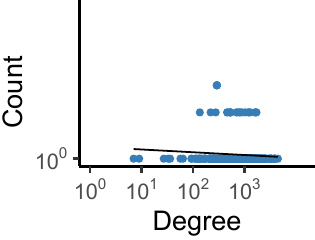} &
			\includegraphics[height=0.762in]{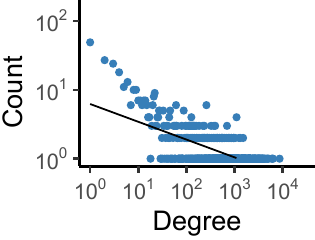} &	
			\includegraphics[height=0.762in]{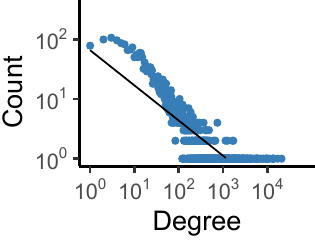} &
			\includegraphics[height=0.762in]{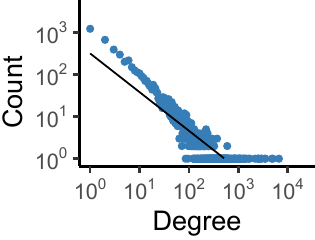} &
			\includegraphics[height=0.762in]{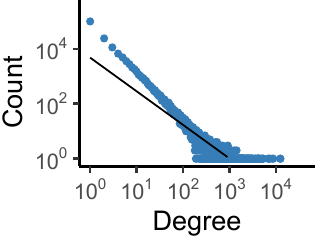} &
			\includegraphics[height=0.762in]{coauth_deg}
			\\ 
			\midrule
			\rotatebox[origin=l]{90}{\small $ $ $ $ Edge Sizes}
			&
			\includegraphics[height=0.762in]{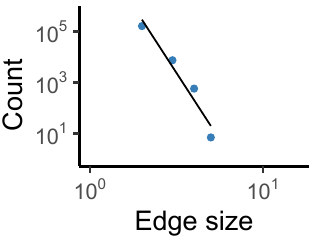} &
			\includegraphics[height=0.762in]{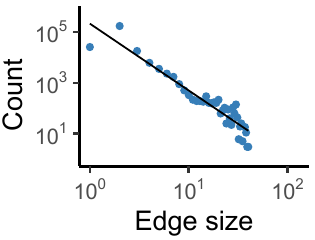} &	
			\includegraphics[height=0.762in]{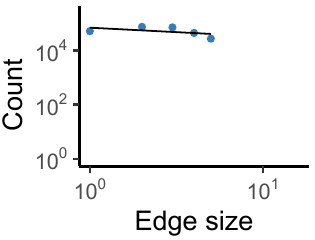} &
			\includegraphics[height=0.762in]{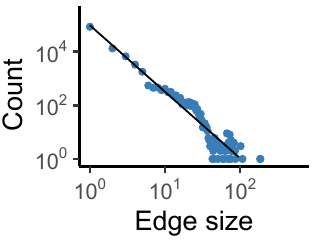} &
			\includegraphics[height=0.762in]{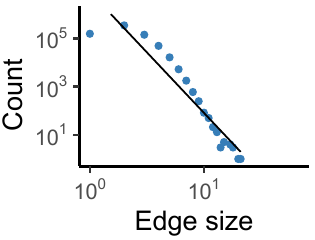} &
			\includegraphics[height=0.762in]{coauth_edge}
			\\
			\midrule
			\rotatebox[origin=l]{90}{\footnotesize Intersection Sizes}
			&
			\includegraphics[height=0.762in]{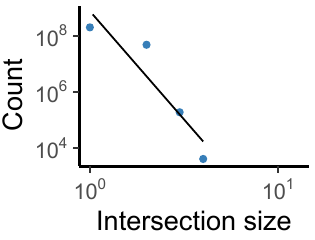} &
			\includegraphics[height=0.762in]{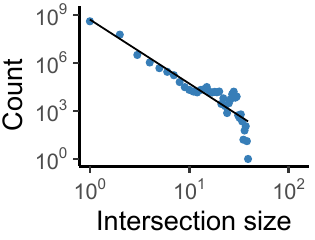} &	 	
			\includegraphics[height=0.762in]{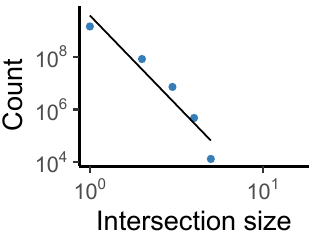} &
			\includegraphics[height=0.762in]{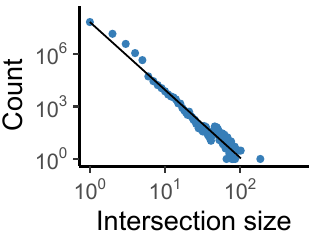} &
			\includegraphics[height=0.762in]{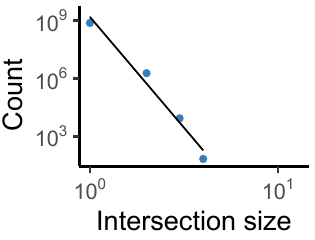} &
			\includegraphics[height=0.762in]{coauth_soi}
			\\
			\midrule
			\rotatebox[origin=l]{90}{\small Singular Values}
			&
			\includegraphics[height=0.762in]{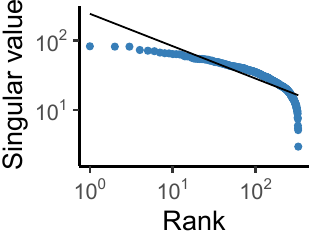} &
			\includegraphics[height=0.762in]{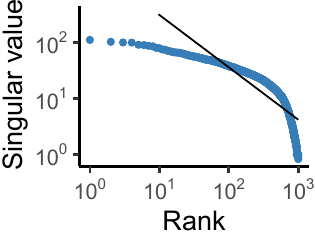} &	 
			\includegraphics[height=0.762in]{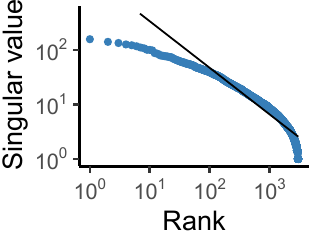} &
			\includegraphics[height=0.762in]{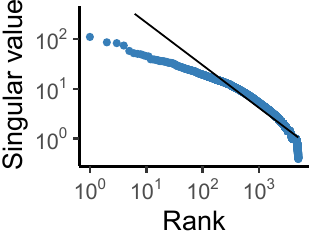} &
			\includegraphics[height=0.762in]{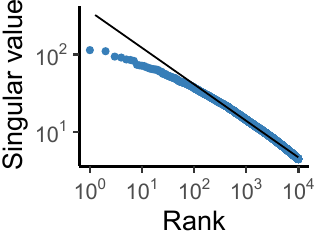} &
			\includegraphics[height=0.762in]{coauth_sv}
			\\
			\bottomrule
		\end{tabular}
	}
	\vspace{-5mm}
\end{figure*}

In this section, we introduce some basic notations and important concepts used throughout the paper.

\smallsection{Hypergraph.}
\label{sec:prelims:hypergraph}
A \emph{hypergraph} $G=(V, E)$ consists of a set $V$ of \emph{nodes} and a set $E\subseteq 2^V$ of \emph{hyperedges}.
Each hyperedge $e$ is a subset of $V$, and we define the \emph{size} of a hyperedge $e$ as the number $|e|$ of nodes in the hyperedge. 
Note that conventional graphs are a special case of hypergraphs where the sizes of all hyperedges are two.
The \emph{degree} of a node $v$, denoted by $deg(v)$, is defined as the number of hyperedges containing $v$.
The \emph{neighborhood} of a node $v$, denoted by $N(v)$, is defined as the set of \emph{neighbors}, each of which is contained together with $v$ in one or more hyperedges.

\smallsection{Incidence Matrix.}
\label{sec:prelims:incidence}
The \emph{incidence matrix} $I \in \{0, 1\}^{|V|\times |E|}$ of a graph $G=(V, E)$ indicates the membership of the nodes $V$ in the hyperedges $E$.
Each $(i,j)$-th entry $I_{ij}$ of $I$ is $1$ if and only if the $j^{th}$ hyperedge in $E$ contains the $i^{th}$ node in $V$.

\smallsection{Effective Diameter.}
\label{sec:prelims:effDiam}
A \emph{path} in a hypergraph is a sequence of hyperedges in which any two immediate hyperedges have a non-empty intersection and the \emph{length} of the path is the length of the sequence.
The \emph{distance} between two nodes is defined as the length of a shortest path in which the first and the last hyperedges include one of the two nodes and the other, respectively. 
The \emph{diameter} of the hypergraph is a maximum distance between any pairs of nodes.
Since any disconnected nodes in a hypergraph makes the diameter infinite, we consider the \emph{effective diameter} \cite{leskovec2007graph}, which is defined as the smallest $d$ such that the paths of length at most $d$ connect $90\%$ of all reachable pairs of nodes.\footnote{Linear interpolation is used to find such $d$.}

\smallsection{Heavy-tailed Distribution.}
\label{sec:prelims:heavytail}
Tails of \emph{heavy-tailed distributions} decay slower than exponential distributions (i.e., they are not exponentially bounded).
Power law distributions are typical examples of the heavy-tailed distributions.
For instance, a discrete power law distribution $P$ of a random variable $X$ satisfies, possibly in a limited range, the relationship $P(X) \propto 1/|X|^{\alpha}$ for some constant $\alpha > 0$.  
Note that this relationship appears as a straight line on the log-log plot of the probability distribution over the range of the random variable.

\smallsection{Goodness of Fit.}
\label{sec:prelims:fit}
It is difficult to argue that an empirical dataset genuinely follows a target probability distribution, since other statistical models unexamined yet may describe the dataset with a better fit.
Thus, it is more sound to rely on comparative tests which compare the goodness of fit of candidate distributions \cite{alstott2014powerlaw,clauset2009power}.
We especially utilize the log likelihood-ratio test \cite{woolf1957log, mclachlan1987bootstrapping} to this end.
When a dataset $\Dcal$ with two candidate distributions $A$ and $B$ is given, the test computes $\log \left( \frac{\Lcal_A(\Dcal)}{\Lcal_B(\Dcal)} \right)$, where $\Lcal_A(\Dcal)$ stands for the likelihood of the dataset $\Dcal$ with respect to the candidate distribution $A$.
Positive ratios imply the distribution $A$ is a better description available for the dataset between the two distributions, and negative ratios imply the opposite.

\smallsection{Hypergraph Sequence.}
\label{sec:prelims:formulation}
Allowing for multiple hyperedges created at the same time, we use $e_t^s$ to denote the set of hyperedges created at time $t$ and $e_t$ to denote a hyperedge in $e_t^s$. 
Given a sequence $\edgesseq$ of a set of time-stamped hyperedges for some $T > 0$, we define a sequence $\{G_t = (V_t, E_t)\}_{t=1}^T$ of hypergraphs evolving under the hyperedge sequence, where the nodes $V_t$ of $G_t$ is $\bigcup_{i=1}^{t}\bigcup e_i^s$ and the hyperedges $E_t$ of $G_t$ is $\bigcup_{i=1}^t e_i^s$.
While the final snapshot $G_T$ of the hypergraph sequence is of our interest in exploring structural patterns, subsequences of the hypergraph sequence are examined for the understanding of dynamical patterns.

\smallsection{Null Model.}
\label{sec:prelimns:nullModel}
We make use of a null model as a counterpart of an evolving hypergraph $\graphseq$ so as to emphasize the significance of examined patterns in real-world hypergraphs in Section~\ref{sec:patterns}.
Formally, the null model is constructed from a sequence of a set of hyperedges $\nulledgesseq$ such that there is a one-to-one correspondence between $\edges$ and $\nulledges$.
Specifically, $\nulledge$ in $\nulledges$ has the same size with its corresponding hyperedge $\edge$ in $\edges$, while $\nulledge$ consists of randomly selected $|\nulledge|$ nodes from $V_t$.
Note that this null model has the same distribution of hyperedge sizes with its corresponding dataset.

\begin{table}[t]
	\centering
	\caption{\label{tab:dataset} 
		The real-world hypergraphs used in our study.
	}
	\scalebox{1}{
		\begin{tabular}{l|l|l|l}
			\toprule
			Dataset & \# of Nodes & \# of Hyperedges & Summary  \\ 
			\midrule
			\midrule
			\contact & $327$		  &  $172,035$  	&  Social Interaction \\ 
			\email 			& $1,005$ 	  &  $235,263$  	&  Email \\
			\tags	& $3,029$ 	  &  $271,233$  	&  Q\&A \\
			\ndc  	& $5,556$ 	  &  $112,919$		&  Drug \\ %
			\thread 	& $176,445$   &  $719,792$ 		&  Q\&A \\
			\coauth		& $1,924,991$ &  $3,700,067$ 	&  Coauthorship \\ %
			\bottomrule
		\end{tabular}
	}
	\vspace{-6mm}
\end{table}

\section{Structural and Dynamical Patterns}
\label{sec:patterns}

\begin{table*}[t]
	\centering
	\caption{\label{table:static:test}
		\but{Log-likelihood ratio of two competing distributions: three heavy-tailed dists. versus exponential dist.}.
		We consider power law \textbf{(pw)}, truncated power law \textbf{(tpw)}, and log-normal \textbf{(logn)} distributions as candidates for the heavy-tailed distributions.
		We report the log-likelihood ratio normalized by its standard deviation and make the largest one among the three candidates boldfaced.
		We also provide the $p$-value of the boldfaced one if it is larger than $0.05$, where the $p$-value is the significance value for the acceptance of the heavy-tailed distribution involved in the ratio.
		As suggested in Figure~\ref{fig:static:patterns}, the first three patterns generally have the best fit to (truncated) power law distributions,
		while singular values are best described by log-normal distributions.
		Remark that \model also shows the same trend.
	}
	\scalebox{0.96}{
		\hspace{-3mm}
		\begin{tabular}{cx{0.92cm}x{0.9cm}x{0.92cm}x{0.92cm}x{0.9cm}x{0.9cm}x{0.9cm}x{0.9cm}x{0.9cm}x{0.9cm}x{0.9cm}x{0.9cm}}
			\toprule
			\multicolumn{1}{c}{} &
			\multicolumn{3}{c}{Degrees}  &
			\multicolumn{3}{c}{Hyperedge Sizes${}^{*}$} &
			\multicolumn{3}{c}{Intersection Sizes} &
			\multicolumn{3}{c}{Singular Values} 
			\\
			\cmidrule(lr){2-4}
			\cmidrule(lr){5-7}
			\cmidrule(lr){8-10}
			\cmidrule(lr){11-13}
			\multicolumn{1}{c}{Heavy-tailed Dist.} &
			pw &
			tpw &
			logn &
			pw &
			tpw &
			logn &
			pw &
			tpw &
			logn &
			pw &
			tpw &
			logn 
			\\
			\midrule
			\midrule
			\contact  
			& $-0.612$ & $\textbf{0.495}^{\dagger}$ & $-0.011$
			& -  & -  &  -  
			& -  & -  &  -  
			& $-290$ & $-205$ & $\textbf{116}$
			\\
			\email  
			& $0.013$ & $\textbf{2.01}$  & $1.72$
			& $28.6$ & $\textbf{34.0}$ & $32.8$  
			& $454$ & $\textbf{463}$ & $461$
			& $-219$ & $-157$ & $\textbf{75.7}$  
			\\
			\tags  
			& $8.60$  & $\textbf{9.51}$  &  $9.45$
			& $-713$  & $-111$  &  $\textbf{103}$
			& -  & -  &  -  
			& $-407$  & $-285$  &  $\textbf{145}$
			\\
			\ndc
			& $3.69$ & $\textbf{3.90}$ & $3.83$ 
			& $29.5$ & $\textbf{31.8}$ & $30.2$ 
			& $39.1$ & $\textbf{39.9}$ & $\textbf{40.0}$
			& $-361$ & $-249$ & $\textbf{121}$
			\\
			\thread  
			& $38.0$ & $\textbf{38.5}$ & $38.4$
			& $0.786$ & $\textbf{1.04}^{\ddagger}$ & $1.02$
			& -  & -  &  -  
			& $-1171$ & $-832$ & $\textbf{445}$
			\\
			\coauth  
			& $187$ & $\textbf{206}$ & $204$
			& $4.14$ & $4.14$ & $\textbf{4.15}$
			& $2.28$ & $\textbf{2.36}$ & $\textbf{2.36}$
			& $-661$ & $-471$ & $\textbf{268}$
			\\
			\midrule
			{\footnotesize \textbf{\model (Proposed)}}
			& $19.6$  & $\textbf{27.0}$  & $25.9$  
			& $-0.737$  & $\textbf{1.36}^{\mathparagraph}$  &  $1.29$  
			& -  & -  &  -  
			& $-525$  & $-364$  &  $\textbf{210}$ 
			\\
			\bottomrule
			\multicolumn{13}{l}{${}^{\ast}$ \contactl, \tags, \thread, and \model have a small range of hyperedge sizes and intersection sizes, which make some ratio not available.} \\
			\multicolumn{13}{l}{${}^\dagger$ The $p$-value is $0.62$. ${}^\ddagger$ The $p$-value is $0.30$. ${}^{\mathparagraph}$ The $p$-value is $0.17$.	
		}
		\end{tabular}
	}
	\vspace{-5mm}
\end{table*}

In this section, we examine characteristics of real-world hypergraphs prevalent across $6$ distinct domains and shed light on common four structural and three dynamical patterns at the macroscopic level.
We summarize some basic statistics on the datasets \cite{benson2018simplicial, Mastrandrea-2015-contact, leskovec2007graph} in Table~\ref{tab:dataset} and briefly describe the characteristics of each dataset below.
\begin{itemize}[leftmargin=*]
	\item \textbf{contact-high-school (\contact):} each node is a student, and each hyperedge is a set of individuals interacting each other as a group during an unit interval.
	\item \textbf{email-Eu (\email):} each node is an email address at an European research institution, and each hyperedge consists of the sender and all recipients of an e-mail.
	\item \textbf{tags-ask-ubuntu (\tags):} \url{askubuntu.com} is a question-and-answer website, where one can ask a question with up to $5$ tags attached.
	Each node and hyperedge correspond to a tag and the set of tags attached to a question, respectively.
	\item \textbf{NDC-substances (\ndc):} each node is a substance, and each hyperedge indicates the set of substances which a drug is made of.
	\item \textbf{threads-math-sx (threads):} \url{math.stackexchange.com} is a question-and-answer website.
	Each node is a user on the website, and each hyperedge corresponds to the set of users participating in a thread that lasts for at most $24$ hours.
	\item \textbf{coauth-DBLP (\coauth):} each node is an author, and each hyperedge corresponds to the set of authors in a publication recorded on DBLP.
\end{itemize}

We first observe the heavy-tailed distributions of degrees, hyperedge sizes, intersection sizes, and singular values of incidence matrices as the structural patterns.
Then, we elaborate on diminishing overlaps of hyperedges, densification, and decreasing diameter as the dynamical patterns.

\subsection{Structural Patterns}
\label{sec:patterns:static}

\begin{figure}[t]
	\vspace{-2mm}
	\centering
	\hspace{-2mm}
	\subfigure[Degrees]{\label{fig:static:null:degree}
		\includegraphics[width=.3\columnwidth]{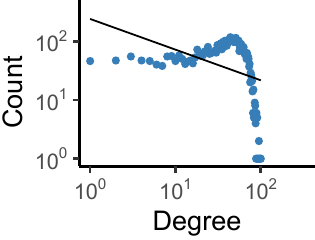}
	} %
	\subfigure[Size of Intersections]{\label{fig:static:null:intersize}
		\includegraphics[width=.3\columnwidth]{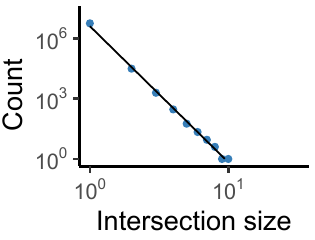}
	} %
	\subfigure[Singular Values]{\label{fig:static:null:sing}
		\includegraphics[width=.3\columnwidth]{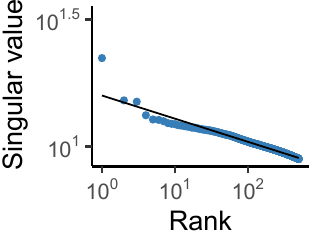}
	} %
	\vspace{-1mm}
	\caption{\label{fig:static:null}
		\but{Comparison with the null model generated from} \but{\emph{substance}}.
		Its bell-shaped degree distribution and the significantly dominant singular value, which appear across all the datasets, are main differences from real hypergraphs.
		Note that the range of hyperedge sizes is significantly reduced compared to that of \ndc.
	}
	\vspace{-6mm}
\end{figure}

In this section, we study four tendencies in the final snapshots of the $6$ real hypergraph sequences and present all the results in Figure~\ref{fig:static:patterns}.
We demonstrate that \textbf{(S1)} degrees, \textbf{(S2)} hyperedge sizes, \textbf{(S3)} intersection sizes, and \textbf{(S4)} singular values of incidence matrices obey heavy-tailed distributions.
Moreover, out of three probable candidates - power law, truncated power law, and log-normal distributions - quantitative validation based on the log-likelihood test supports that the first three are described best by the truncated power law distribution, and the last is closest to the log-normal distribution.
These patterns are peculiar in the sense that the random null model described in Section~\ref{sec:prelimns:nullModel} cannot reproduce them, as illustrated in Figure~\ref{fig:static:null}.

\begin{figure*}[t]
	\centering
	\caption{\label{fig:temporal:patterns}
		\but{Dynamical patterns shown in six real datasets: diminishing overlaps, densification, and shrinking diameter}.
		The slopes smaller than $1$ in the first row indicate decreasing ratios of intersecting pairs.
		The slopes larger than $1$ for the edge density imply increasing average degrees of the hypergraphs.
		The effective diameters eventually decrease.
		For the \emph{\thread} dataset, the effective diameter starts to shrink almost in the end.
		See Section~\ref{sec:patterns:temporal} for details.
	}
	\scalebox{0.98}{
		\hspace{-3mm}
		\begin{tabular}{p{0.02\textwidth}|cccccc}
			\toprule
			{\footnotesize Data} & \contact & \email & \tags & \ndc & \thread & \coauth \\
			\midrule
			\midrule
			\rotatebox[origin=l]{90}{\small Intersecting Pairs}
			&
			\includegraphics[height=0.762in]{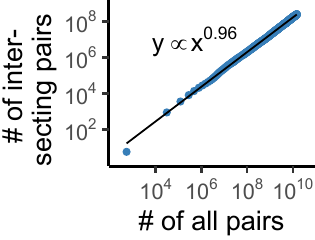} &
			\includegraphics[height=0.762in]{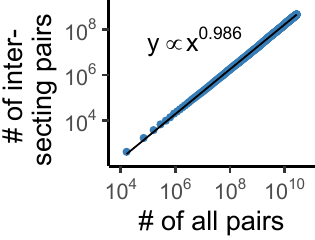} &  	 	
			\includegraphics[height=0.762in]{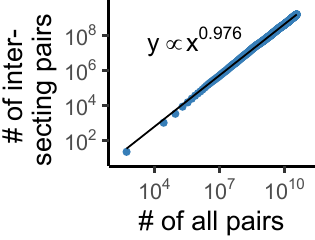} &
			\includegraphics[height=0.762in]{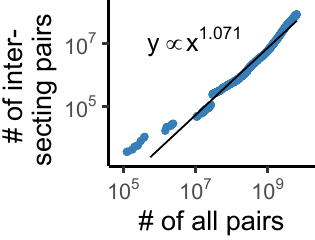} &
			\includegraphics[height=0.762in]{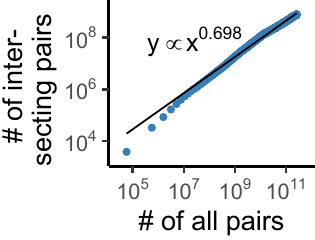} &
			\includegraphics[height=0.762in]{coauth_doi} 
			\\
			\midrule
			\rotatebox[origin=l]{90}{\small Edge Density}
			&
			\includegraphics[height=0.762in]{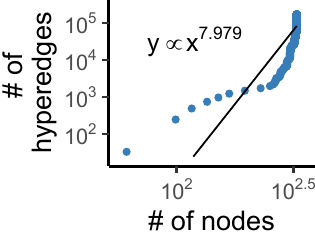} &
			\includegraphics[height=0.762in]{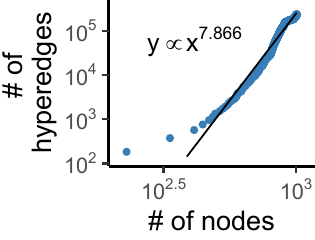} &  	
			\includegraphics[height=0.762in]{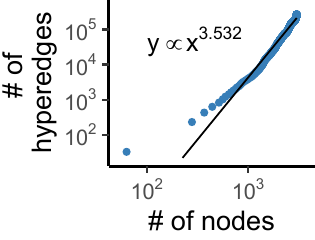} &
			\includegraphics[height=0.762in]{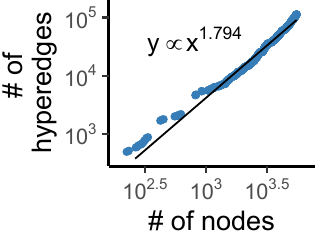} &
			\includegraphics[height=0.762in]{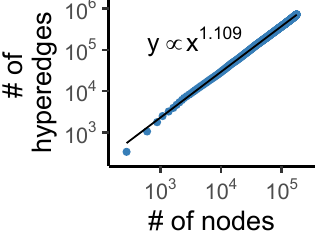} &
			\includegraphics[height=0.762in]{coauth_ed}  
			\\
			\midrule
			\rotatebox[origin=l]{90}{\small \quad Diameter}
			&
			\includegraphics[height=0.762in]{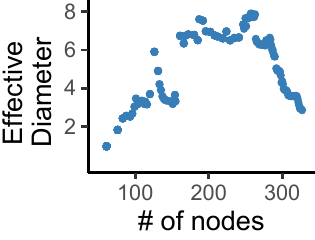} &
			\includegraphics[height=0.762in]{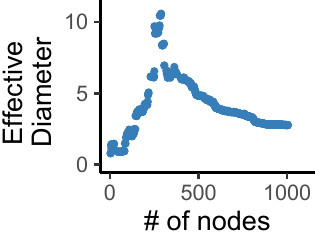} &  
			\includegraphics[height=0.762in]{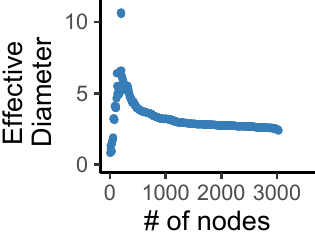} &
			\includegraphics[height=0.762in]{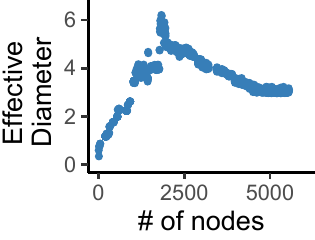} &
			\includegraphics[height=0.762in]{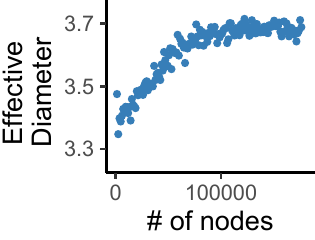} &
			\includegraphics[height=0.762in]{coauth_diam}  
			\\
			\bottomrule
		\end{tabular}
	}
	\vspace{-5mm}
\end{figure*}

\smallsection{S1. Heavy-tailed degree distribution.}
\label{sec:patterns:static:degree}
The degree distributions of real hypergraphs generally fall under the class of \htails.
Note that our investigation of the degree distributions differs from the previous approaches \cite{manh2020multi} where the degrees of nodes $2$-projected graphs are investigated.

As reported in Table~\ref{table:static:test}, the comparative test implies that for each dataset, the three representative \htails have a better fit than an \expdist.
Moreover, the likelihood ratio and the well-fitted straight lines on the log-log scale in Figure~\ref{fig:static:patterns} suggest that for the degree distributions, the (truncated) power law distribution is the best candidate among the \htails.

In Figure~\ref{fig:static:null:degree}, the skewed bell-shaped degree distribution of the random null model also indicates the peculiarity of the observed patterns.

\smallsection{S2. Heavy-tailed hyperedge size distribution.}
\label{sec:patterns:static:edgeSize}
Hyperedge sizes are also found to follow heavy-tailed distributions. 
Note that we use all hyperedges, regardless of their sizes, while only small hyperedges are taken into consideration in \cite{benson2018sequences, manh2020multi}.
The distribution of hyperedge sizes generally shows a better fit to the \htails than to the \expdist, and the similar reasoning based on Figure~\ref{fig:static:patterns} and Table~\ref{table:static:test} implies that the (truncated) power law distribution is the best description available.
Note that we do not report the distribution of hyperedge sizes of the null model, as its definition results in the same distribution with its corresponding real hypergraph.

\smallsection{S3. Heavy-tailed intersection size distribution.}
\label{sec:patterns:static:soi}
The intersection sizes of two hyperedges (i.e., the number of nodes commonly contained in two hyperedges), which offer insights into pairwise intersections of hyperedges, also follow \htails.
As shown in Table~\ref{table:static:test}, the \htails are probable descriptions for this pattern, and the reported likelihood ratios in Table~\ref{table:static:test} together with the straight lines in Figure~\ref{fig:static:patterns} back up the goodness of fit to the (truncated) power law distribution.

In comparison with the null model, whose intersection-size distribution is shown in Figure~\ref{fig:static:null:intersize}, although its overall distribution seems similar, the range of intersection sizes and the frequency of each intersection size are significantly different from those of the corresponding hypergraph.

\smallsection{S4. Skewed singular values.}
\label{sec:patterns:static:sing}
This pattern means that the singular values of the incidence matrices of real hypergraphs are generally heavy-tailed.
As certified in Table~\ref{table:static:test}, the singular-value distribution is best described by the log-normal distribution, which is one of the \htails.
Apparently in Figure~\ref{fig:static:patterns}, the tails of the singular-value distributions decay faster than those of the three distributions in \textbf{(S1-S3)}.

In contrast to the real distribution, we observe in Figure~\ref{fig:static:null:sing} that the null model has a highly dominant singular value.

\subsection{Dynamical Patterns}
\label{sec:patterns:temporal}

Now we move on to the investigation of three dynamical patterns in the six real hypergraphs sequences.
We devise a quantity to measure how much the hyperedges are overlapped in overall and show that \textbf{(T1)} overall overlaps of the hyperedges decrease over time.
Just as phenomena of densification and shrinking diameter are well known to take place in real graphs,
the two dynamical patterns are still prevalent in real hypergraphs; \textbf{(T2)} average degrees increase over time and \textbf{(T3)} effective diameters decrease over time.

\smallsection{T1. Diminishing overlaps.}
\label{sec:patterns:temporal:overlap}
We first define the \emph{density of interactions} as follows to capture the overall overlaps of hyperedges:
for a hypergraph $G_t = (V_t, E_t)$ at time $t$,
\begin{equation}\label{equation:doi}
	DoI(G_t) := \frac{|\{ \{e_i, e_j \} \vert e_i \cap e_j \neq \emptyset \textup{ for } e_i, e_j\in E_t  \}|}
	{|\{ \{e_i, e_j\} \vert e_i, e_j \in E_t \}|}	
\end{equation}
In words, it is simply the ratio of the number of intersecting pairs of hyperedges to the number of all possible pairs of hyperedges, which amounts to $\binom{|E_t|}2$.
Note that this quantity can range from $0$ to $1$. 

Using $y(t)$ and $x(t)$ to indicate the numerator and the denominator, respectively, in Equation~\ref{equation:doi}, we take a closer look at the log-log plot of $y(t)$ over $x(t)$.
As shown in the first row of Figure~\ref{fig:temporal:patterns}, the slopes $s$ of fitted lines on the plots implies the formula $DoI(G_t) = y(t)/x(t) = O(x(t)^{-(1-s)})$.
As the slopes $s$ are usually smaller than $1$, the density of interactions decreases over time.
Decreasing $DoI(G_t)$ indicates that the ratio of overlapping pairs of hyperedges decreases and that the intersections of hyperedges become less frequent in overall.

\smallsection{T2. Densification.}
\label{sec:patterns:temporal:densification}
In the following two subsections, we confirm that dynamical phenomena in real graphs - densification and shrinking diameter established in a seminal work \cite{leskovec2007graph} - still take place in real hypergraphs. 

We proceed with the same manner as in Section~\ref{sec:patterns:temporal:overlap}.
For a hypergraph $G_t = (V_t, E_t)$ at time $t$, we plot $|E_t|$ over $|V_t|$ on the log-log scale in the second row of Figure~\ref{fig:temporal:patterns} and compute the slopes $s$ of the fitted lines, which lead to the formula $|E_t| \propto |V_t|^s$.
The slopes larger than $1$ in general indicate that the average degrees of hypergraphs, formulated as $2|E_t|/|V_t|$, increase over time, and thus the densification is also valid for real hypergraphs.

\smallsection{T3. Shrinking diameter.}
\label{sec:patterns:temporal:shrinking}
In the third row of Figure~\ref{fig:temporal:patterns}, we show how the effective diameters of hypergraphs change over time.
We observe that the effective diameters of all real hypergraphs eventually decrease over time.
In the \emph{\thread} dataset, while the decrease is marginal, the effective diameter starts to decrease almost in the end.

\begin{figure*}[t]
	\centering
	\caption{\label{fig:model} 
		\but{Comparison between a real hypergraph and those generated by \model with distinct parameters}.
		Each row for \model($p, q$) with the burning probability $p$ and expanding probability $q$ shows holistic behaviors of the model.
		With suitable probabilities (e.g., $p=0.51$ and $q=0.2$), \model successfully reproduces all examined patterns.
		The overall comparison indicates larger $p$ and $q$ lead to faster densification and shrinking diameter, respectively.
		See Section~\ref{sec:model} for details.
	}
	\scalebox{0.83}{
		\hspace{-4mm}
		\begin{tabular}{c|c|ccccccc}
			\toprule
			\multicolumn{2}{c}{} &
			\multicolumn{4}{c}{\textsc{Structural Patterns}} &
			\multicolumn{3}{c}{\textsc{Dynamical Patterns}} 
			\\
			\cmidrule(lr){3-6}
			\cmidrule(lr){7-9}
			\multicolumn{2}{c}{}
			& Degrees & Hyperedge sizes & Intersection Sizes & Singular Values & Intersecting Pairs & Edge Density & Diameter
			\\
			\cmidrule(lr){3-6}
			\cmidrule(lr){7-9}
			\multicolumn{1}{c|}{\multirow{4}{*}{\rotatebox[origin=l]{90}{\textbf{\model (Proposed)}
						\qquad}}}&
			\rotatebox[origin=l]{90}{($0.51, 0.2$)} &
			\includegraphics[height=0.762in]{model_051_02_deg} &
			\includegraphics[height=0.762in]{model_051_02_edge} &  	 	 			
			\includegraphics[height=0.762in]{model_051_02_soi} &
			\includegraphics[height=0.762in]{model_051_02_sv} &
			\includegraphics[height=0.762in]{model_051_02_doi} &
			\includegraphics[height=0.762in]{model_051_02_ed} &
			\includegraphics[height=0.762in]{model_051_02_diam}
			\\
			&
			\rotatebox[origin=l]{90}{($0.51, 0.3$)} &
			\includegraphics[height=0.762in]{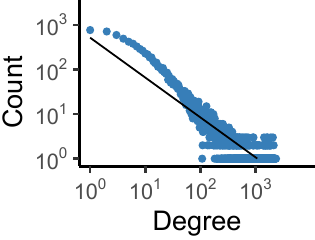} &
			\includegraphics[height=0.762in]{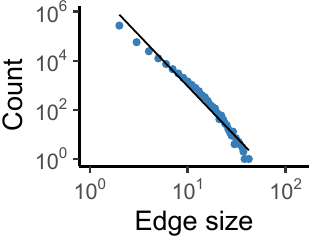} &  	 	 			
			\includegraphics[height=0.762in]{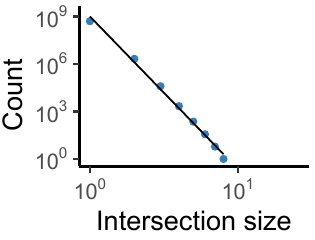} &
			\includegraphics[height=0.762in]{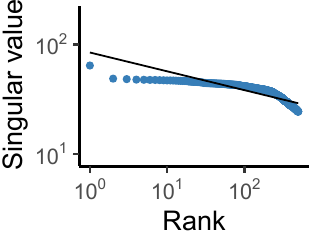} &
			\includegraphics[height=0.762in]{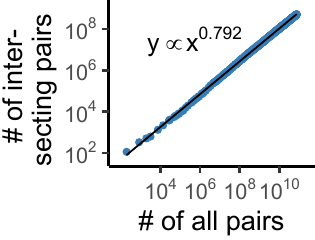} &
			\includegraphics[height=0.762in]{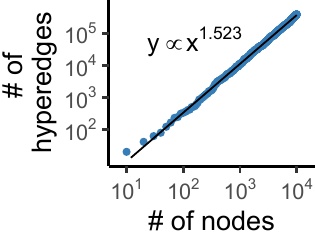} &
			\includegraphics[height=0.762in]{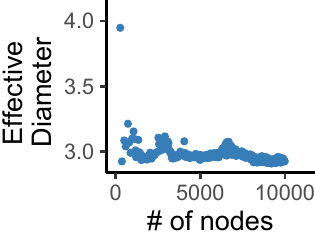}
			\\
			&
			\rotatebox[origin=l]{90}{($0.45, 0.2$)} &
			\includegraphics[height=0.762in]{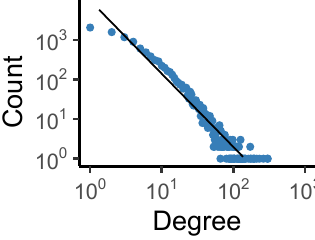} &
			\includegraphics[height=0.762in]{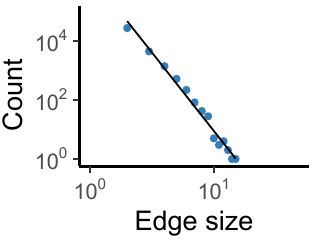} &  	 	 			
			\includegraphics[height=0.762in]{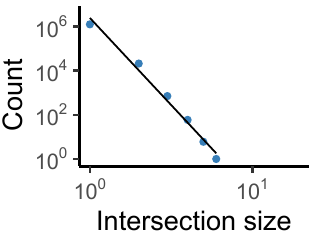} &
			\includegraphics[height=0.762in]{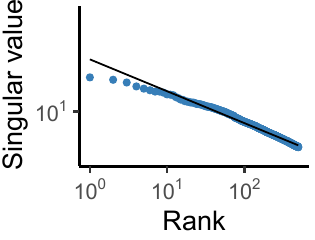} &
			\includegraphics[height=0.762in]{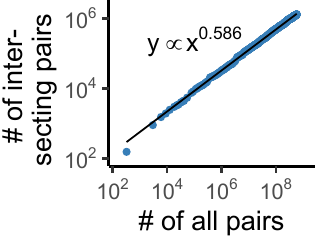} &
			\includegraphics[height=0.762in]{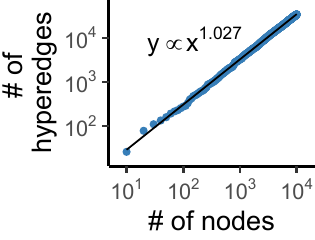} &
			\includegraphics[height=0.762in]{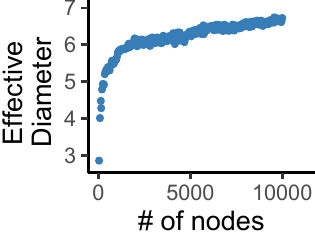}
			\\
			&
			\rotatebox[origin=l]{90}{($0.45, 0.3$)} &
			\includegraphics[height=0.762in]{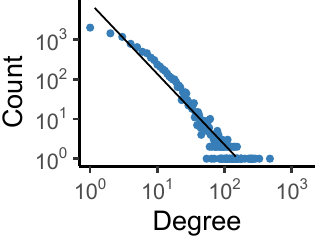} &
			\includegraphics[height=0.762in]{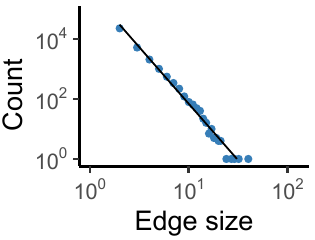} &  	 	 			
			\includegraphics[height=0.762in]{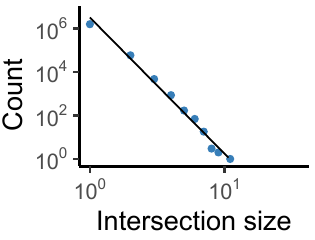} &
			\includegraphics[height=0.762in]{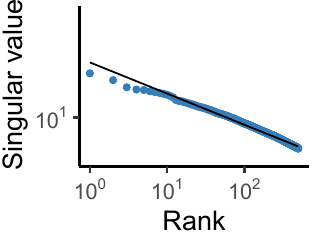} &
			\includegraphics[height=0.762in]{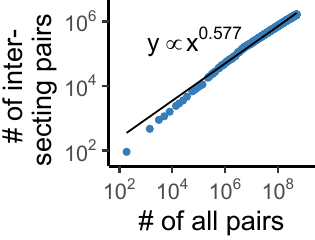} &
			\includegraphics[height=0.762in]{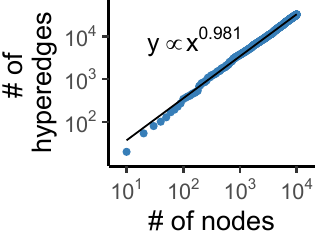} &
			\includegraphics[height=0.762in]{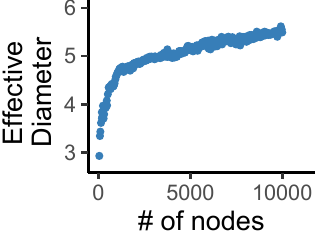}
			\\
			\cmidrule(lr){3-6}
			\cmidrule(lr){7-9}
			\multicolumn{2}{c|}{\rotatebox[origin=l]{90}{\coauth\xspace (Real)}} &
			\includegraphics[height=0.762in]{coauth_deg} &
			\includegraphics[height=0.762in]{coauth_edge} &  	 	 			
			\includegraphics[height=0.762in]{coauth_soi} &
			\includegraphics[height=0.762in]{coauth_sv} &
			\includegraphics[height=0.762in]{coauth_doi} &
			\includegraphics[height=0.762in]{coauth_ed} &
			\includegraphics[height=0.762in]{coauth_diam}
			\\
			\bottomrule
		\end{tabular}
	}
	\vspace{-5mm}
\end{figure*}

It is worthy to note that although shrinking diameter and densification seem intuitively compatible with one another, one can construct counterexamples that have only one of the two dynamical patterns, which are illustrated in the second and third row in Figure~\ref{fig:model}.

\section{Proposed Model: Hypergraph Forest Fire}
\label{sec:model}

In parallel with the establishment of underlying patterns in real hypergraphs,
we propose a growth model \model (\textbf{Hyper}graph \textbf{F}orest \textbf{F}ire) for realistic hypergraph generation.
In a nutshell, for each new node, \model simulates a `forest fire', which is spread stochastically through existing hyperedges, to decide the nodes each of which forms a size-$2$ hyperedge with the new node. Then, \model simulates a forest fire again for each created hyperedge to expand it.

\model reproduces all seven examined patterns and additionally reproduces all five structural patterns found in previous work \cite{manh2020multi} without relying on oracles, i.e., unexplainable outside information.
Notably in \model, the mechanisms on individual nodes are simple and intuitive, while they do not directly impose but eventually lead to the examined patterns.

In Section~\ref{sec:model:descript}, we describe \model in detail with a pseudocode and rationales behind it.
In Section~\ref{sec:model:exp}, we confirm that it successfully reproduces all the seven patterns as well as five additional patterns.
Lastly in Section~\ref{sec:model:analysis}, we explore the parameter space of \model.

\setlength{\textfloatsep}{1pt}%
\begin{algorithm}[t]
	\caption{\label{algo:model} \model: the proposed model for realistic hypergraph generation.	
	}
	\DontPrintSemicolon
	\KwIn{burning probability: $p$, expanding probability: $q$, \\
		\quad\quad\quad timespan (i.e., the target number of nodes): $T$}
	\KwOut{evolving hypergraph: $\graphseq$}
	\SetKwFunction{algo}{Generator}
	\SetKwFunction{proc}{Burning}
	\SetKwProg{myalg}{Algorithm}{}{}
	\myalg{\algo{$p, q, T$}}{
		$G_0 \setgets$ hypergraph with $1$ node and $0$ hyperedges\;
		Initialize $tie$ measuring how close two nodes are\; 
		\ForEach{time $t$ in $[1, \dots, T]$}{
			$\Vt \setgets V_{t-1} \cup$ $\{$new node $u\}$ \& $\edges \setgets \{\}$\;
			$w \setgets $ random ambassador from $V_{t-1}$\;
			$Burned \setgets$ \proc($w, p$)\;
			\ForEach{$v$ in $Burned$}{
				Increase $tie(\{u, v\})$ by $1$\;
				$Burned \setgets$ \proc(v, q)\;
				$\edges \addgets$ hyperedge $Burned\cup\{u\}$\;
			}
			
			$E_t \setgets E_{t-1} \cup e_t^s$\;
		}
		\Return{$\graphseq$}\;}{}
	\nonl $ $ \newline
	\setcounter{AlgoLine}{0}
	\SetKwProg{myproc}{Subroutine}{}{}
	\myproc{\proc{source, prob}}{
		$Burned = \{ \}$ \& $Queue \setgets $ empty queue \;
		$Queue \addgets source $\;
		\While{$Queue \neq \emptyset$}{
			$s \setgets$ node popped from $Queue$\;
			$Burned \addgets s $\;
			$n \sim$ geometric dist. with mean $\frac{prob}{1-prob}$\;
			$Candidates \setgets N(s)\setminus Burned\setminus Queue$\;
			$Queue \addgets$ $n$ neighbors in $Candidates$ in decreasing order of $tie(\{s, neighbor\})$\;
		}
		\Return{$Burned$}
	}
\end{algorithm}

\subsection{Description of model}
\label{sec:model:descript}

We provide the pseudocode of \model in Algorithm~\ref{algo:model}.
Note that, except the target number of nodes (i.e., $T$), it needs only two parameters: \emph{burning probability} $p$ and \emph{expanding probability} $q$.
\model starts with a hypergraph with one node and no hyperedge; and it initializes tie strengths, which measure the closeness of two nodes, to $0$. Then, it repeats the following steps for each new node $u$.
\begin{enumerate}
	\item[(1)] The new node $u$ chooses a random ambassador $w$ from the hypergraph so far and burns the ambassador.
	\item[(2)] Burn $n$ neighbors of the ambassador $w$ in descending order of tie strength, where  $n$ is sampled from the geometric distribution with mean $p/(1-p)$.
	In case of burning a part of neighbors with the same tie strength, proceed with random selection.
	\item[(3)] Recursively apply (2) to each burned neighbor by viewing a burned neighbor as a new ambassador of the new node $u$.
	Nodes cannot be burned twice until the recursion ends.
	\item[(4)] For each burned node $v$, 
	form a hyperedge $\{u, v\}$ and increase tie strength of $\{u, v\}$ by $1$.%
	\item[(5)] For each hyperedge created in (4),
	reset the burning history and
	 start the burning process at the burned node $v$ in which we use the geometric distribution with mean $q/(1-q)$, expand the hyperedge until the process ends.
\end{enumerate}
Each step has a straightforward rationale, and it may be well understood through an example of a coauthorship network.
Suppose a student joins a research community, advised by a supervisor (ambassador).
When introducing colleagues to the student for coworking, the supervisor is more likely to introduce intimate peers (considering tie strength).
Those who cowork with the student recursively introduce their close peers for follow-up research (recursive burning). 
When a group of researchers looks for a new researcher to work with (hyperedge expansion), a referrer in the group is likely to introduce those who the referrer has worked with many times before than a totally new researcher (considering tie strength).

\smallsection{Comparison with \cite{leskovec2007graph}.}
Our model \model has several differences from the forest fire model \cite{leskovec2007graph}, which \model is inspired by.
Most importantly, the forest fire model generates conventional graphs instead of hypergraphs.
Moreover, it relies on the `orientation' of burning, requiring two parameters: forward and backward burning probabilities.
Without relying on the orientation, it fails to achieve both shrinking diameters and power law distributions of degrees.
On the other hand, \model achieves both by additionally taking tie strength (instead of the orientation) into account, without requiring an additional parameter.
As a result, \model successfully copes with the complexity due to hyperedges of any size, while 
maintaining the number of parameters to two (i.e., one probability for neighbor selection and the other for expansion).

\smallsection{Remarks.}
\model suggests possible local dynamics on individual nodes of hypergraphs that give rise to the \emph{macroscopic} patterns examined in this paper.
Thus, predicting an order of hyperedges at the \emph{microscopic} level goes out of scope.
Also, for simplicity, it does not have a particular parameter for repetition of hyperedges.
Constructing a model tackling these sides is left for future work.

\begin{figure*}[t]
	\centering
	\caption{\label{fig:manh}
		\but{Compatibility with previous work {\cite{manh2020multi}}}.
		\model and a real dataset have similar heavy-tailed degree distributions at each decomposition level, and they even have the same tendency of deviating from the fitted lines at higher decomposition levels.
		Singular values of \model and the dataset basically lie on their fitted lines, regardless of decomposition levels. 
		See Section~\ref{sec:model:exp} for details.
	}
	\scalebox{0.95}{
		\begin{tabular}{c|cccccc}
			\toprule
			\multicolumn{1}{c}{} &
			\multicolumn{3}{c}{Degrees}  &
			\multicolumn{3}{c}{Singular Values} 
			\\
			\cmidrule(lr){2-4}
			\cmidrule(lr){5-7}
			\multicolumn{1}{c}{Level}
			& Node & Edge & Triangle
			& Node & Edge & Triangle
			\\
			\cmidrule(lr){1-4}
			\cmidrule(lr){5-7}
			\rotatebox[origin=l]{90}{\textbf{\model}}
			\rotatebox[origin=l]{90}{\textbf{(Proposed)}} & 
			\includegraphics[height=0.762in]{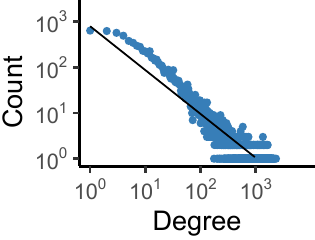} &
			\includegraphics[height=0.762in]{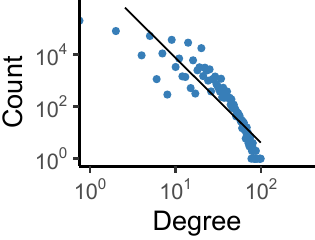} & 
			\includegraphics[height=0.762in]{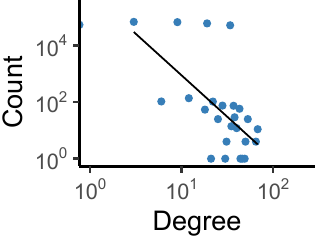} &
			\includegraphics[height=0.762in]{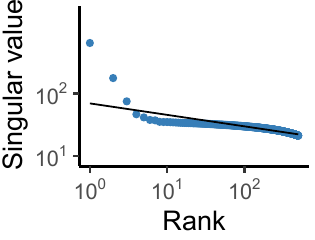} &
			\includegraphics[height=0.762in]{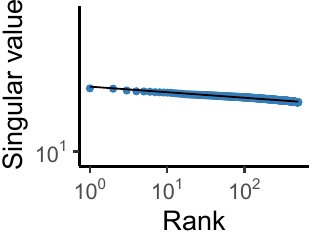} & 
			\includegraphics[height=0.762in]{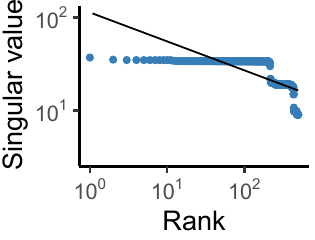}
			\\
			\rotatebox[origin=l]{90}{\ndc} 
			\rotatebox[origin=l]{90}{\quad (Real)} & 
			\includegraphics[height=0.762in]{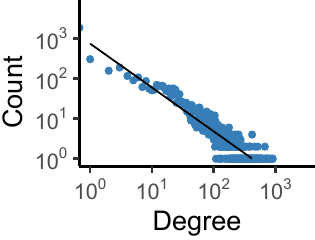} &
			\includegraphics[height=0.762in]{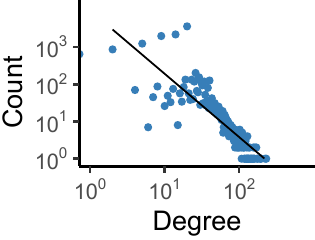} &
			\includegraphics[height=0.762in]{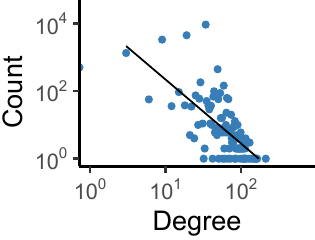} &
			\includegraphics[height=0.762in]{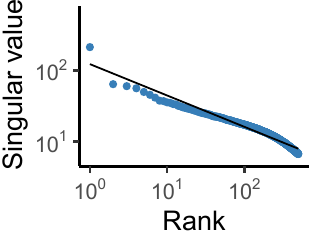} &
			\includegraphics[height=0.762in]{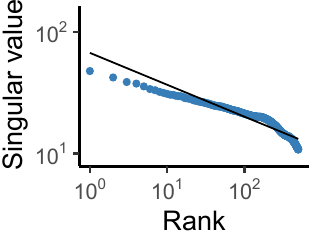} & 
			\includegraphics[height=0.762in]{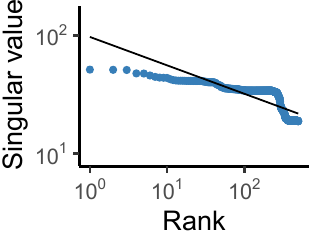}
			\\\bottomrule
		\end{tabular}
	}
	\vspace{-5mm}
\end{figure*}

\subsection{Experimental results}
\label{sec:model:exp}

We examine the proposed model \model thoroughly, in terms of the established patterns, as in Section~\ref{sec:patterns}.
We also examine its $n$-level decomposed graphs, as suggested in \cite{manh2020multi}.

\smallsection{Observed patterns.}
We demonstrate that \model with suitable parameter values\footnote{We investigate a synthetic hypergraph with $10,000$ nodes and set the burning probability to $0.51$ and the expanding probability to $0.2$.} reproduces four structural and three dynamical patterns investigated in Section~\ref{sec:patterns}.

In Figure~\ref{fig:model}, we provide overall patterns of models with four distinct parameter settings.
The model in the first row apparently exhibits all the patterns consistent with the representative results of one representative real dataset.
Especially for the four structural patterns, including the distribution of degrees, hyperedge sizes, intersection sizes, and singular values,
we confirm in Table~\ref{table:static:test} that all the distributions of the model have the same tendency with real hypergraphs.
The four distributions are best suited to \htails, and among probable \htails, the best description for each pattern is also consistent with that of real hypergraphs; the first three and the last are close to the (truncated) power law distributions and log-normal distribution, respectively.

We also check that the model achieves the three dynamical patterns - diminishing overlap, densifying graph, and decreasing diameter - in Figure~\ref{fig:model}.

\smallsection{Structural patterns in decomposed graphs.}
The multi-level decomposition method \cite{manh2020multi} provides a way of reducing hypergraphs to $m$ ordinary graphs for the largest size $m$ of hyperedges.
Each reduced graph is referred to as an $n$-level decomposed graph,\footnote{Formally, 
the $\mathit{n}$\textit{-level decomposed graph} of a hypergraph $G=(V, E)$ is defined as $G_{(n)}=(V_{(n)}, E_{(n)})$ where
\begin{align*}
& V_{(n)}  := \{v_{(n)} \in 2^{V} : |v_{(n)}| = n \text{ and } \exists e\in E \text{ s.t. } v_{(n)} \subseteq e \}, \\
& {E_{(n)}  := \{\{u_{(n)},v_{(n)}\} \in \binom{V_{(n)}}{2} : \exists e\in E \text{ s.t. } u_{(n)} \cup v_{(n)} \subseteq e \}.}
\end{align*}
} which focuses on the interplays between pairs of size-$n$ subsets of nodes.
In \cite{manh2020multi}, it is shown that the decomposed graphs of real hypergraphs generally retain five structural patterns of real networks - giant connected components, heavy-tailed degree distributions, small diameters, high clustering coefficients, and approximately heavy-tailed singular values of adjacency matrices - unlike the random null model that we adopt in Section~\ref{sec:prelim}.
Then, the structural patterns are suggested as judging criteria for real hypergraphs.

To make the setting consistent with \cite{manh2020multi}, we delete all hyperedges of size larger than $5$ and consider the decomposition level up to $4$, where each level is named as the \emph{node} level, the \emph{edge} level, the \emph{triangle} level, and the $4$\emph{clique} level, respectively.
Moreover, we focus on the node level especially for effective diameter and clustering coefficient as highlighted in \cite{manh2020multi}, and up to the triangle level for the distributions of degree and singular values due to the lack of space.

We compare in Figure~\ref{fig:manh} the overall tendency of the degree and singular-value distributions of the decomposed graphs of \model with that of real hypergraphs.
The degree distributions of decomposed graphs, regardless of whether they are real or synthetic, seem to follow a power law, while the plotted points tend to deviate from a fitted line as the decomposition level increases.
The singular-value distributions also reveal similar trends in both synthetic and real datasets.
In Table~\ref{table:manh}, we report the effective diameter and clustering coefficient at the node level, and the ratio of the size of the largest connected component to the total size of each decomposed graph. 
Similar to the real datasets, at the node level, the decomposed graph from \model also attains a small effective diameter and high clustering coefficient.
The level at which the largest connected component becomes small varies across the datasets, and our model maintains a giant connected component up to the edge level.

\begin{table}[t]
	\centering
	\small
	\caption{\label{table:manh} 
		Size of the largest connected component at each decomposition level, and effective diameter and clustering coefficient at the node level.
		The red colors indicate the level at which decomposed graphs no longer retain a giant connected component.
		Note that \model also shows a similar trend with real datasets;
		its giant connected component starts to shatter after the edge level, its effective diameter is small, and it has similar clustering coefficient with the real datasets.
	}
	\scalebox{0.85}{
		\begin{tabular}{ccccccc}
			\toprule
			\multicolumn{1}{c}{\multirow{2}{*}{}} &
			\multicolumn{4}{c}{Largest Connected Component}  &
			\multicolumn{1}{c}{\multirow{2}{*}{Diam.}} &
			\multicolumn{1}{c}{\multirow{2}{*}{Clus. Coeff.}}\\
			\cmidrule{2-5}
			& Node & Edge & Triangle & $4$clique &  						 &
			\\
			\midrule
			\contact  
			& $1.00$  & $0.46$  & $\red{0.02}$ & $\red{0.02}$ & $2.63$ & $0.50$
			\\
			\email
			& $0.98$ & $0.70$ & $0.80$ & $0.41$ & $2.80$ & $0.49$
			\\
			\tags
			& $0.997$ & $0.94$ & $0.71$ & $\red{0.22}$ & $2.41$ & $0.61$
			\\
			\ndc
			& $0.58$ & $0.78$ & $0.35$ & $\red{0.02}$ & $3.56$ & $0.42$
			\\
			\thread
			& $0.87$ & $0.45$ & $\red{0.03}$ & $\red{0.0004}$ & $3.68$ & $0.37$
			\\
			\coauth
			& $0.86$ & $0.53$ & $\red{0.05}$ & $\red{0.0006}$ & $6.84$ & $0.60$
			\\
			\midrule
			\textbf{\model}
			& $1.00$ & $0.52$ & $\red{0.0003}$ & $\red{0.0005}$ & $3.00$ & $0.69$
			\\
			\bottomrule
		\end{tabular}
	}
\end{table}

\subsection{Exploring parameter space}
\label{sec:model:analysis}

In Figure~\ref{fig:model}, we present qualitative patterns in \model with parameters set from all the combinations of two burning probability and two expanding probability.
We observe that the burning probability and the expanding probability are mainly concerned with how fast a hypergraph densifies and its effective diameter decreases, respectively.
Note that with low burning probabilities, the effective diameter of a hypergraph is actually increasing in spite of the densification.

\section{Conclusions}
\label{sec:conclusions}

Despite the omnipresence of hypergraphs, relatively little attention has been paid to structural and dynamical patterns of real-world hypergraphs.
Toward more extensive and thorough understanding of the real-world hypergraphs, we closely examine four structural and three dynamical patterns prevalent in them.
The former includes the heavy-tailed distributions of \textbf{(S1)} degrees, \textbf{(S2)} hyperedge sizes, \textbf{(S3)} intersection sizes, and \textbf{(S4)} singular values of incidence matrices.
The latter includes \textbf{(T1)} diminishing overlaps, \textbf{(T2)} densification, and \textbf{(T3)} shrinking diameter.
We validate the significance of these patterns by comparison with a null model and statistical tests.

We also propose a generative model \model, which captures all seven observed patterns and also shows results compatible with previous findings.
Especially, it has only two scalars (e.g., burning and expanding probabilities) as parameters, and it does not rely on any external information for imitating realistic patterns. 
Surprisingly, \model is made up of simple and intuitive mechanisms on individual nodes, which non-trivially lead to all the examined macroscopic patterns.

For reproducibility, we make the source code and datasets used in this work publicly available at \url{https://git.io/JUfAr}.

\vspace{1mm} 
{\small \smallsection{Acknowledgements} 
This work was supported by National Research Foundation of Korea (NRF) grant funded by the
Korea government (MSIT) (No. NRF-2020R1C1C1008296) and Institute of Information \& Communications
Technology Planning \& Evaluation (IITP) grant funded by the Korea government (MSIT) (No. 2019-0-00075, Artificial Intelligence Graduate School Program (KAIST)).}

\bibliographystyle{IEEEtran}
\bibliography{BIB/yunbum}

\end{document}